# Impact of organic capping layer on the magnetic anisotropy of ultrathin Co films

L. Gladczuk, K. Lasek, R. Puzniak, M. Sawicki, P. Aleshkevych, W. Paszkowicz, R. Minikayev, I. N. Demchenko, Y. Syryanyy, and P. Przyslupski

*Institute of Physics, Polish Academy of Sciences, Aleja Lotnikow 32/46, PL-02668 Warsaw, Poland.*

Email: gladl@ifpan.edu.pl

**Abstract**

Using combined magnetometry and ferromagnetic resonance studies, it is shown that introduction of a hydrocarbon cover layer leads to an increase of the contribution of the surface anisotropy to the effective magnetic anisotropy energy of hexagonal close-packed 0001 cobalt films, largely enhancing the perpendicular anisotropy. Due to a weak electronic interaction of the organic molecules with the Co atoms, an increase of surface anisotropy could be explained by the presence of an electric field at the organic-material/Co interface and by modification of surface atoms charge configuration.

## I. Introduction

Application of organic materials in spintronic devices has recently received a great deal of attention [1]. Long spin-relaxation lifetimes, hyperfine interactions, and complex couplings in organic materials lead to a very large magnetoresistance, as it has been demonstrated for the vertical spin valves (SV) utilizing organic spacers, such as tris (8-hydroxyquinoline) aluminum (Alq3) or rubrene, for instance [2,3,4,5,6,7]. Moreover, the discovery of electronic interactions between a molecular organic layer and a cobalt electrode allowed for the demonstration of magnetoresistive behavior even with a single magnetic electrode [7]. It is known that, for low thicknesses of cobalt films facing nonmagnetic layers, the large interface energy overcomes the volume and the shape anisotropy energy – the generally regarded two major contributions to magnetic anisotropy energy (MAE) – and the easy magnetization axis is directed out of the plane [8]. Materials with perpendicular magnetic anisotropy (PMA) are particularly valuable in the development of perpendicular magnetic tunneling junctions (p-MTJ), used in perpendicular spin-transfer torque magnetic random access memories (p-STT MRAMs) [9]. The tunneling properties and the stability of the magnetic tunneling junctions strongly depend on the structural and magnetic arrangement of the electrodes and barrier material. Several studies demonstrated an influence of nonmagnetic metals (NM) cover layers on the interfacial PMA of Co [10,11,12,13]. It has been recognized, that the effect is determined by the orbital moment anisotropy of Co and could be modified by an interface hybridization of the NM layer (e.g., its *d* or *p* orbitals) with Co 3*d* orbitals [14,15,16,17].

The influence of organic materials on PMA is expected to be small due to their weak spin-orbit coupling; however, recent works show that the out-of-plane MAE could also be enhanced, e.g., at a graphene/Co interface [18,19,20,21]. The enhancement of the surface energy is attributed to the incorporation of π bonds of organic molecules with 3*d* electrons of Co.

Organic memory cells with perpendicular anisotropy are especially attractive due to size and power density reduction. Although numerous experiments have been reported for various molecular spin-valves [5,22], very little is known about the effect of organic-material/ferromagnetic interface on the magnetic anisotropy of those structures. New possibilities arose with the discovery of the spin-dependent hybridization at the ferromagnet/molecule interface, offering new perspectives to tailor spintronic devices properties [23,24,25,26,27]. However, no quantitative measurements have been reported until now, and the understanding of the role of organic overlayers on the MAE is still lacking. Studies of physical phenomena occurring at the organic/inorganic interface are expected to lead to considerable progress in the development of the organic spintronics technology, including memories, sensors and new processing elements, as well as conceptually different future applications.

In this paper, the role of an organic cover layer on the magnetic anisotropy of ultrathin Co films is presented. It is shown that the interaction between the organic molecules and the metal surface atoms enhances the surface energy and induces perpendicular anisotropy in the magnetic film.

## II. Experimental

Heterostructures of Mo/Au/Co were deposited by molecular beam epitaxy technique on 11$\bar{2}$0 sapphire substrates. A detailed description of sample preparation and crystallographic characterization is presented in Refs. [28,29]. Results described in this paper were obtained for the samples with the Co layer thickness *d* varying from 1.8 to 5 nm. Two sets of samples were fabricated. For the first one the deposition was terminated after obtaining the required thickness of Co layer, and then the sample was immersed into a beaker with a vacuum oil (APIEZON "B" Oil; manufactured by Metropolitan-Vickers Electrical Co. Ltd.) located in the loading chamber. These samples are marked Au/Co/H-c since the hydrocarbon (H-c) layer with the thickness of the order of 1 μm was introduced on the top of Au/Co structure. For the second set (reference samples) a gold protective layer (10 nm) was deposited and these samples are labeled as Au/Co/Au.

The X-ray diffraction patterns were obtained in a powder diffractometer (Cu $K_\alpha$ radiation) equipped with a graphite monochromator allowing for the collection of diffraction data of high accuracy (see Ref. [30] for details). The magnetic properties of the samples were measured with an MPMS XL Quantum Design superconducting quantum interference device (SQUID) magnetometer with a use of long Si strips to facilitate supporting of the sample in the sample chamber [31]. Ferromagnetic resonance measurements were performed at room temperature with a conventional X-band ($f$ = 9.38 GHz) Bruker EMX spectrometer. For these measurements, the samples were mounted on a quartz sample holder, and the FMR spectra were recorded for various angles of the external magnetic field ($\mathbf{H_{ext}}$) with respect to the sample plane. The core-level spectra were determined by X-ray photoelectron spectroscopy (XPS) using monochromated Al $K_\alpha$ radiation (h$\nu$ = 1486.6 eV), from an area of 5×0.1 mm$^2$. The XPS measurements were carried out at room temperature in *normal emission* set-up; the electrons take-off angle with respect to the sample surface was 90 deg. For those measurements, the organic coverage from the sample surface was removed by dipping sample in acetone before it was loaded into the vacuum chamber, in order to increase the XPS signal intensity. The metallic cobalt sample was sputtered using the argon ion source operating with the 1 kV beam energy and the incident Ar-ion beam axis of 35 deg to the sample surface plane. To avoid the surface damage of the Co/H-c sample, Ar-ion etching was not performed prior to XPS measurements. The binding energy scale was calibrated with a tolerance of ±0.05 eV, applying the Au $4f_{7/2}$ (84 eV) line.

## III. Results

### A. X-ray diffraction

The X-ray diffraction patterns of Au/Co/Au and Au/Co/H-c structures with Co layer thickness of 5 nm (as an example) are presented in Fig. 1. In agreement with previously published data [28], the three strongest peaks in the $2\theta$ range from 74 to 106 deg correspond to the 11$\bar{2}$0 sapphire substrate, 220 Mo peak, and 222 Au peak underlayers, respectively. A small variation in the intensity of the reflected beam is visible, which is related to the absorption of the X-rays by the organic protective layer. Pronounced satellite peaks are present for all deposited layers in the vicinity of 220 Mo and 222 Au peaks, indicating a good crystallographic order both of these layers and the Mo/Au interface. The low-intensity broad structure at $2\theta \cong 97$ deg corresponds to the 0004 Co diffraction peak from hexagonal phase yielding the $c$ value of 4.114±0.001 Å for Au/Co/Au and 4.118±0.001 Å for Au/Co/H-c. The $c$ lattice parameter for 5 nm Co film structure is of about 1% larger than that for the bulk hcp-Co crystal equal to 4.0611 Å. The identity of lattice parameters for the structures with both types of cover layers excludes difference in the impact of lattice mismatch on the magnetic anisotropy of Au/Co/Au and Au/Co/H-c. The 1% difference between the $c$-lattice constant of both Au/Co/Au and Au/Co/H-c structures and that of bulk hcp-Co crystal could be explained by the incoherent growth of (0001) oriented hcp Co lattice structure on (111) Au substrate [32].

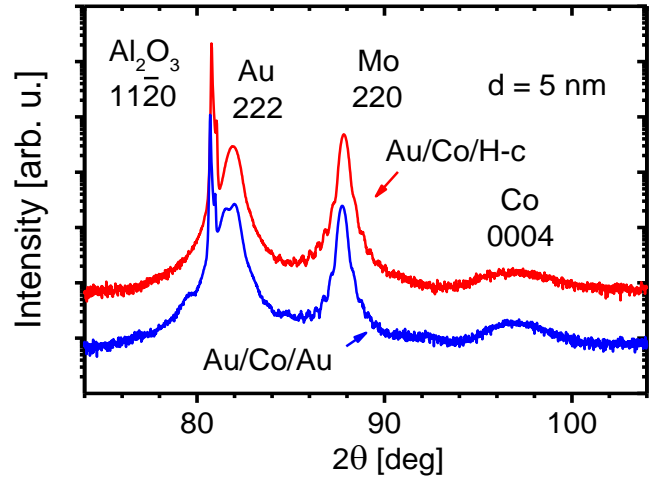

**Figure 1**. X-ray diffraction patterns of Au/Co/Au and Au/Co/H-c structures with 5 nm Co layer thickness. Both patterns are vertically shifted for clarity.

### B. Magnetometry

Magnetic hysteresis loops, $M(H)$, measured at 300 K and 10 K for two Au/Co/H-c samples with $d$ of 1.8 and 2.7 nm are shown in Fig. 2. Only the easy direction of magnetization is presented, which is found to be the easy plane for the 2.7 nm and uniaxial – along the $c$ axis – for the 1.8 nm cobalt layer. These orientations are identical at both temperatures of 300 K and

10 K, and the results are temperature-independent at the full temperature range studied. Despite the fact that for the 10 K measurements both layers have been cooled down at 0.5 T, the recorded $M(H)$ are symmetrical with respect to $H = 0$ line. The absence of the exchange bias effect for Au/Co/H-c samples, otherwise seen in unprotected layers, indicates a high efficiency of the Co surface protection against oxidation by the organic layer studied here. This observation, in turn, indicates that organic compounds used in the daily basis in the laboratories may serve as an efficient yet economical mean of sample (surface) protection for long-term storages.

The magnetization of Au/Co/H-c heterostructures with Co thickness of 1.8 nm saturates at about 20 mT (at $T = 300$ K) and 200 mT (at $T = 10$ K), indicating that the fields of that magnitude are sufficient to reorient all spins along the easy axis, directed perpendicularly to the sample plane. Furthermore, it is found, that the thickness of the Co layer also determines the coercivity, which is approximately three times larger in the 1.8 nm thick Au/Co/H-c layer (exhibiting PMA) than in the 2.7 nm one (the easy plane case). Interestingly, this relationship is seen at the whole studied temperature range, as exemplified in Fig. 2 for 300 and 10 K, despite of the coercivity field variation from 10 mT at 300 K to 150 mT at 10 K for easy axis of 1.8 nm thick film and from 2.5 mT at 300 K to 57 mT at 10 K for easy plane of 2.7 nm thick film.

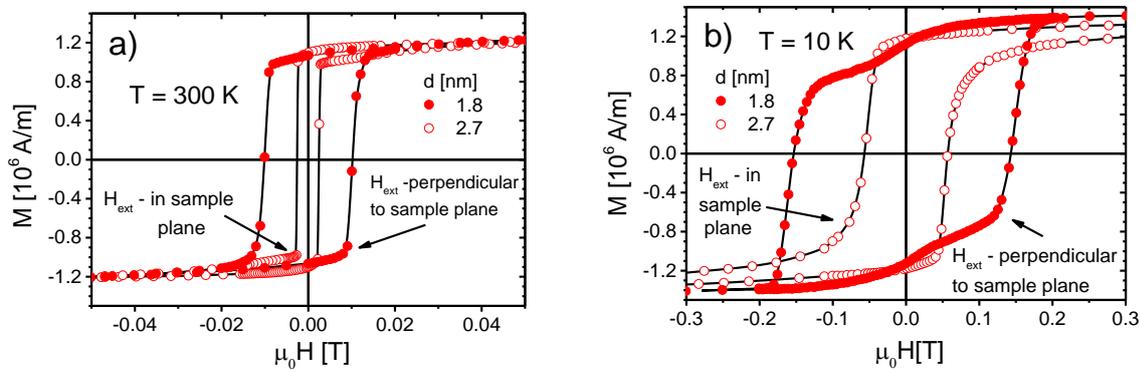

**Figure 2.** Magnetic hysteresis of 1.8 nm (full symbols) and 2.7 nm (open symbols) thick 0001 Co on 111 Au samples at temperatures: a) 300 K and b) 10 K. Both were covered by an organic cover layer of vacuum oil. Only an easy axis behavior is shown: this axis is perpendicular to the sample plane for 1.8 nm layer and an in-plane orientation for the 2.7 nm one.

### C. Ferromagnetic resonance

The out-of-plane angular dependence of FMR resonance field at room temperature as a function of the angle $\theta_H$ between applied magnetic field and normal to the surface for Au/Co/H-c and Au/Co/Au - reference structures, is presented in Fig. 3a and 3b. The evolution of the resonance field as a function of $\theta_H$ for the samples thicker than 2.4 nm (Fig. 3a) indicates that the easy axis of magnetization is oriented in the plane of the sample, as expected for thick Co films with relatively small surface anisotropy in comparison to the shape anisotropy term. For Au/Co/H-c and Au/Co/Au heterostructures with Co thickness close to 2.2 nm, the magnitude of the perpendicular resonance field is comparable to the in-plane resonance field, $H_{res}$, indicating the significance of the fourth-order contribution to the magnetocrystalline anisotropy.

For the samples thinner than 2.2 nm, the minimum and maximum value of $H_{res}$ appear for perpendicular ($\theta_H = 0$ deg) and parallel to the sample surface ($\theta_H = 90$ deg) directions of the external magnetic field, respectively. The minimum value of $\mu_0 H_{res}(\theta_H)$ of about 40 mT for 1.8 nm Au/Co/H-c layer is significantly lower than that for Au/Co/Au ($\mu_0 H_{res}$ = 133 mT), indicating a higher surface anisotropy energy for Co/H-c interface in comparison to Co/Au interface. The increase of the magnitude of $H_{res}$, measured along the cobalt c-axis, with the increase of Co layer thickness indicates an enhancement of the in-plane shape anisotropy.

The influence of the thickness of the Co layer on the magnetic anisotropy in the evolution of $H_{res}(\theta_H)$ is clearly visible: the extremum value of the $H_{res}$ at $\theta_H = 0$ deg for the thinner sample is significantly lower in comparison to that of thicker one (see Fig 3 b).

Phenomenologically, the expression for the magnetic free energy density of a thin hexagonal layer of Co can be described as [28,32]:

$$E = -\mu_0 M_S H \cos(\theta - \theta_H) + K_{eff}(d)\cos^2\theta + K_4 \cos^4\theta. \quad (1)$$

Here, the first term is the magnetic potential energy of the cobalt layer of saturation magnetization $M_S$, when it is oriented at an angle $\theta$ from the normal to the layer as the result of the application of the external magnetic field $H$ at the angle $\theta_H$ to the z-axis directed perpendicularly to the film plane. The angle-independent coefficient of the second term

$$K_{eff}(d) = K_2 - \frac{K_S}{d} + \frac{\mu_0 M_S^2}{2} \quad (2)$$

is an effective uniaxial component consisting of the first order uniaxial term $K_2$ (bulk), surface anisotropy $K_S$ brought about by the finite thickness of the magnetic layer $d$, and the usual shape contribution. The term $K_4 \cos^4\theta$ in Eq. (1) describes the second order anisotropy contribution [32].

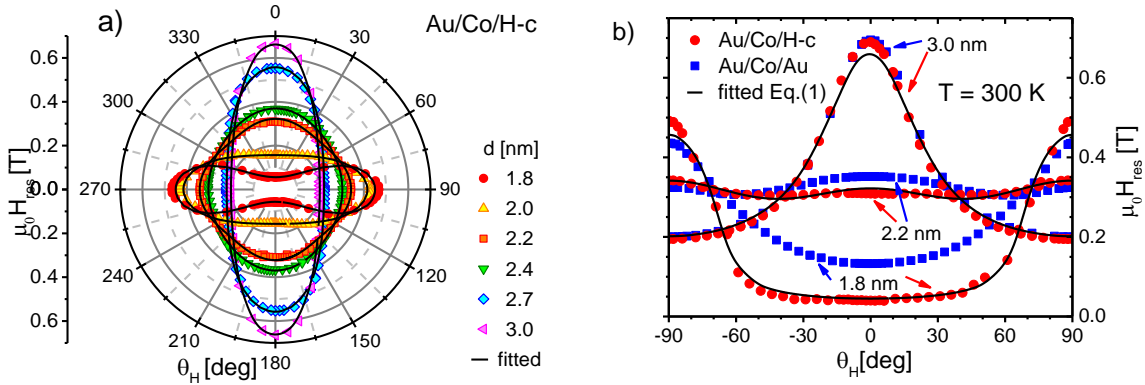

**Figure 3**. The angular dependence of the resonance field $\mu_0 H_{res}$ at room temperature for: a) Au/Co/H-c heterostructures with the Co layer thickness from 1.8 to 3.0 nm, b) two type Au/Co/H-c and Au/Co/Au (reference) heterostructures – selected thicknesses of Co layer. The solid lines are the best fits of Eq. 1 with fixed $M_S = 1.2 \times 10^6$ A/m.

We have fitted Eq. (1) to experimental angular dependences of the resonance field (see Fig. 3). In the numerical calculations we assumed the saturation magnetization of Co, $M_S = 1.20 \times 10^6$ A/m, as was determined from SQUID measurements, independent of Co layer thickness, leaving $K_{eff}$ and $K_4$, as the only fitting parameters. It turned out that the fidelity of the fitting depended only marginally on $K_4$, whose values varied in the range 0.05–0.20 MJ/m$^3$ without showing any systematic relation to Co layer thickness. Therefore, for the final fitting, the $K_4$ value was fixed at 0.10 MJ/m$^3$, a value corresponding very well to the previously established one for similar Au/Co/Au structure [28], leaving just $K_{eff}$ as the only fitting parameter. The $K_{eff}$ values determined as a function of Co layer thickness are presented in Fig. 4.

It is found that the $K_{eff}$ in both types of structures decreases linearly with $1/d$, what allows by fitting a linear regression base on Eq. (2) to estimate two $d$-independent constants characterizing these structures. The Co bulk anisotropy constant $K_2$ was determined by subtraction of $\mu_0 M_S^2/2$ from y-axis intersection point, and the surface anisotropy $K_S$ specific to each of the structures is the slope of the fitted line.
The latter contains contributions from both, the bottom Au/Co and the top (either Co/Au or Co/H-c) interfaces. Assuming that the bottom Au/Co interfaces are identical in both sets of samples, we set up the room temperature Au/Co interface anisotropy constant as 0.87±0.05 mJ/m$^2$, the value comparable to that obtained previously: 0.83±0.01 mJ/m$^2$ [ 33 ]. Consecutively, the surface anisotropy constant for the Co/H-c interface is 1.08±0.05 mJ/m$^2$. According to our knowledge, it is the highest value of the surface anisotropy constant reported for any cobalt interface. It is obvious that the difference in magnitude of $K_S$ is a result of a different capping layer. The obtained value of anisotropy constant $K_2$ = -0.20±0.05 MJ/m$^3$ is the same for both Au/Co/H-c and Au/Co/Au structures and is higher than that known for bulk hcp cobalt, value: $K_2$ = -0.45×10$^6$ MJ/m$^3$ [34], and may result from magnetostrictive effects. Strain in Co layer could arise from the large lattice mismatch of Au and Co, can affect for the magnitude of $K_2$.

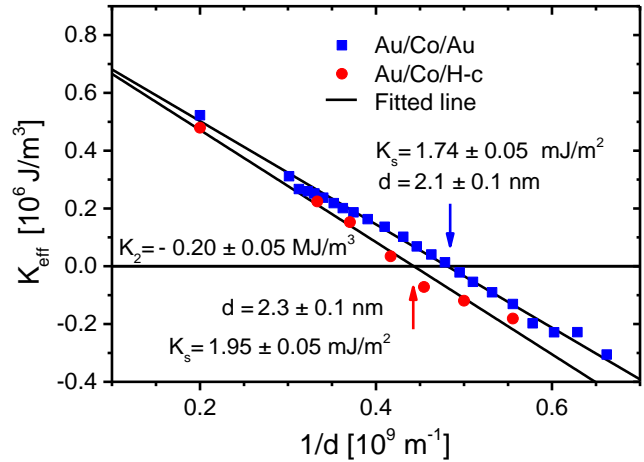

**Figure 4**. The effective anisotropy constant $K_{eff}$ as a function of the inverse Co film thickness $d$, determined by fitting Eq. (1) to the experimental data for all studied samples. Solid lines in Fig. 4 represent the best fit of Eq. (2) to the data points.

### D. X-ray photoelectron spectroscopy

To understand the microscopic origin of the anisotropy evolution associated with Co/H-c interface, the XPS measurements were performed. Comparison of Co 2$p$ high-resolution XPS spectra for metallic cobalt film (before and after argon etching) and Co/H-c sample is shown in Fig. 5. The background was subtracted from the experimental data, and the intensity of the spectrum was normalized to the maximum of Co 2$p_{3/2}$ resonance line. The spectrum corresponding to sputtered cobalt film (red-bottom line in Fig. 5) represents a pure metallic phase of cobalt with 2$p$ spin-orbit splitting of 14.97 eV [35]. The estimated binding energy of Co 2$p_{3/2}$ peak is 778.0±0.15 eV. A distribution of unfilled one-electron levels (conduction electrons) which are available for shake-up like events following core electron photoemission is evident and manifests itself in an asymmetric peak shape of the 2*p* resonances. Naturally oxidized cobalt film (black-top line in Fig. 5) represents the Co 2*p* states of two phases: metallic cobalt and cobalt bonded to oxygen. The observed *chemical shift* is an effective indicator of the charge transfer between oxygen and cobalt states. The binding energy of CoO 2$p_{3/2}$ peak obtained through deconvolution of XPS spectrum is 780.2±0.15 eV, whereas the spin-orbit splitting of 2*p* doublet for cobalt monoxide is about 15.2 eV. Estimated values of energy agree well with those measured for CoO [35, 36]. Moreover, the spectrum demonstrates, along with Co 2*p* photoelectron lines, additional structures in higher binding energies as a result of multiplet splitting (*s*) – this feature arises when an atom contains unpaired electrons (for Co$^{2+}$ ions there are three of them).

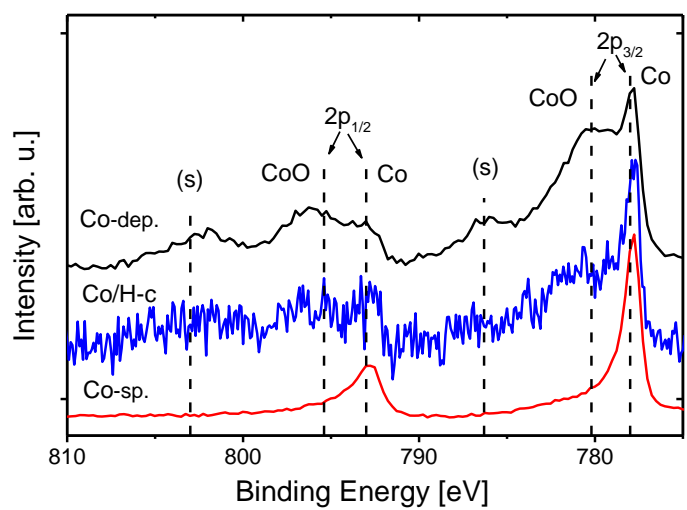

**Figure 5**. Co 2*p* XPS spectra for the Co/H-c sample (middle curve) and Co film (before and after argon ion etching: black-top and red-bottom curves, respectively).

The comparison of metallic cobalt spectra (before and after ion argon etching) to the spectrum obtained for Co/H-c sample (blue-middle line in Fig. 5) reveals a minor contribution of cobalt atoms bonded to oxygen. The Co peaks are shifted towards larger binding energies with respect to cobalt metallic 2*p* peaks due to the oxygen environment, introduced due to a reduction of the thickness of the protective oil layer – a process necessary to permit the XPS investigations (the transfer of sample was performed at atmospheric conditions). Nevertheless, the spectra clearly demonstrate that the Co/H-c interface composition is changed from metallic Co to insulating CoO during the oxidation process, showing that the Co ions located at the Co/H-c interface possess metallic Co–Co and ionic Co–O bonds. There are no other contributions that can be identified in the Co 2*p* XPS spectra.

## IV. Discussion

The main impact of this study is a clear demonstration that the surface anisotropy energy of a magnetic layer can be significantly enhanced by the presence of a layer of organic molecules, for example, a standard vacuum oil. It also turned out that such a layer provides a very effective protection against oxidation over a time scale of at least one year – the duration of this project, most likely - much longer. In fact, we had not recorded any changes of the micromagnetic properties of our very thin layers when the measurements were performed right after the growth and a month later. The identical measurements performed a year later revealed some small deviations, however small comparable to those recorded on the unprotected counterparts. The minuscule changes seen for the protected layers are most probably due to a partial oxidation of the Co layer, an effect of a residual oxygen diffusion through the organic layer.

The high value of surface anisotropy energy ($K_S = 1.08\pm0.05$ mJ/m$^2$) obtained for the Co/H-c interface indicates strong interactions between the cobalt atoms and the organic molecules. The same values of $K_2$ equal to -0.2$\pm$0.05 MJ/m$^3$ for Au/Co/H-c and Au/Co/Au structures indicate comparable magnetic interaction of Co volume atoms in the films (Figs. 3b and 4) for both studied structures and proves that enhancement of surface anisotropy $K_S$ is directly related to the replacement of the gold top layer by the dielectric layer. However, the identification of the specific type of molecule(s) responsible for observed phenomena renders impossible as the ultrahigh vacuum oil is composed of several types of aromatic hydrocarbons of different chain length, and exceeds the scope of this paper.

Previously published results of theoretical and experimental studies attributed the large surface energy of the Au/Co interface to the hybridization of the interfacial Co 3$d$ orbitals and their interaction with the Au 5$d$ orbital. [14,15,37]. Other theoretical investigations, as those in Ref. [17], identified the spin-orbit interaction and the position of energy level all the 3$d$-electron orbitals with respect to the Fermi energy ($E_F$) as the main source of the surface energy enhancement in Co films. The surface energy is significantly strengthened if the energy of Fermi level has a value between the energies associated either with the $d_{xy}$ and $d_{x^2-y^2}$, or with $d_{xz}$ and $d_{yz}$ orbitals [13]. Moreover, the crystal field of the bulk Co atoms strongly affects the energy difference between the surface out-of-plane and in-plane orbitals [14]. Thus, when the crystal field locates the $d_{xy}$ and $d_{x^2-y^2}$ states near to $E_F$ with one being below and the other above the $E_F$ with the energy separation smaller than that in the bulk Co, the perpendicular orbital moment is reasonably enhanced, and the magnitude of the surface energy increases.

The XPS investigation (Fig. 5) reveals that the shift of the Co peak energy (from 778.0 eV to 780.2 eV) is due to the presence of Co–O bonds at the interface, caused by a partial surface oxidation. It was experimentally proved that the presence of oxygen at the Co surface generates the surface energy of 0.35 mJ/m$^2$ [37], which is much lower than in the case when the Co layer is covered by Au. This phenomenon can be explained in terms of the charge transfer between Co and O, which reduces the energy of Co $d$ orbitals pointing toward O (that is out of plane), and creates a splitting between in-plane $d_{xy}$, $d_{x^2-y^2}$ and out-of-plane $d_{xz}$, $d_{yz}$, $d_{z^2}$ orbitals [14,38]. According to the XPS studies, there is no coupling between the constituent molecular metal orbital's and the organic layer and so the observed surface energy enhancement cannot be attributed to the charge transfer effect.

However it has been well established, that the charge screening pushes the energy levels of the molecular orbitals, they are drawn closer to the Fermi level, and a small binding energy in the proximity of the Co surface forms an interfacial dipole with the positive charge at the dielectric side[2]. As a result, the interaction between the molecular film and the metal is increased. The interface dipoles present at Co/H-c interface operate, similarly to the interface orbitals hybridization, as an effective uniaxial crystal field acting on the $d$ orbitals in the first monolayer of Co at the interface [18,25]. This electric field has a significant influence on the interelectronic interactions and has a critical impact on the magnetic anisotropy [14].

Importantly, absence of the localized bonds at the Co/H-c interface guarantees lack of the strain in the metal film and so eliminates the strain origin of changes in the magnetic anisotropy.

Presented above results demonstrate that, by the replacement of the transition metal atoms with carbon-hydrate molecules, the magnetic behavior and, in particular, the magnetic anisotropy can be significantly modified by the electrical field at the interface. Future detailed investigations are necessary to elaborate the accurate nature of the Co/H-c interface states in grater details.

## V. Conclusions

The magnetic properties of the cobalt/hydrocarbon interface were investigated by magnetometry and ferromagnetic resonance techniques. It was demonstrated that the surface energy of Co film could be significantly increased at the cobalt/organic interface and direct an equilibrium magnetization perpendicular to the film plane. Presumably, the redistribution of the charge at the interface, induced by the interaction between the hydrocarbon frontier orbitals and the Co $3d$ band, forms strong dipoles at the metal film surface, which affect the charge configuration of the surface atoms and significantly changes the magnitude of the surface energy. The application of organic materials such as hydrocarbons could provide a platform for the construction of new type spintronic devices. Moreover, organic compounds used in the daily basis in the laboratories may serve as an efficient sample/surface protection for long-term storages.